\documentclass[aps,prl,twocolumn,groupedaddress,amsmath,nofootinbib]{revtex4-1}
\usepackage{graphicx}% Include figure files
\usepackage{subfigure}
\usepackage{mathtools}
\usepackage{mathrsfs} 
\usepackage{dcolumn}% Align table columns on decimal point
\usepackage{bm}% bold math
\usepackage{amsmath}
\usepackage{amsthm}
\usepackage{color}
\usepackage{verbatim}
\usepackage{natbib} 
\usepackage{verbatim}
\usepackage{physics}
\usepackage{amsfonts}
\usepackage{lipsum}
\usepackage{epstopdf}
\usepackage{dsfont}

\usepackage[colorlinks,
            hypertexnames=false,
            linkcolor=blue,
            anchorcolor=blue,
            citecolor=blue,
urlcolor=blue
            ]{hyperref}

\begin{document}
\title{Experimental Observation of Anomalous Complementary Weak Values from Correlated Pairwise Two-State Vectors}

\author{Qian Xie$^{1}$}
\author{Liang Xu$^{1}$}
\email{liangxu.ceas@nju.edu.cn}
\author{Lijian Zhang$^{1}$}
\email{lijian.zhang@nju.edu.cn}
\affiliation{$^1$ National Laboratory of Solid State Microstructures, Key Laboratory of Intelligent Optical Sensing and Manipulation, College of Engineering and Applied Sciences, and Collaborative Innovation Center of Advanced Microstructures, Nanjing University, Nanjing 210093, China\\}

\date{\today}

%\keywords{}
\begin{abstract}
Weak values (WVs) arise from weak measurements performed within a time-symmetric formulation of quantum mechanics, where a system is both pre- and post-selected. Anomalous WVs that lie far outside the eigenvalue spectrum of the observable hold both fundamental and practical significance. However, their generation typically relies on near-orthogonal pre- and post-selection, which confines them to a single post-selection outcome with extremely low success probability. This constraint limits experimental accessibility and hinders the full exploitation of time symmetry. To overcome these limitations, we utilize quantum entanglement and post-selection-controlled operations to generate correlated pairwise two-state vectors. By changing the role of post-selection from passive filtering to active engineering, this approach enables the observation of anomalous complementary WVs associated with mutually exclusive post-selection branches. Our results extend the operational accessibility of time-symmetric quantum structures associated with the two-state vector formalism, and open new avenues for exploring the applications of time symmetry in quantum information processing.
\end{abstract}
\maketitle

\emph{Introduction}.--In the time-symmetric two-state vector formalism (TSVF), a quantum system (QS) at an intermediate time is described by a pair of boundary conditions: a forward-evolving state from an initial pre-selection and a backward-evolving state from a final post-selection~\cite{Aharonov_JPA_1991,Reznik_PhysRevA_1995,Aharonov_time_2010}. Within this framework, weak measurements enable the extraction of partial information from the QS with minimal disturbance, yielding weak values (WVs) to capture the combined influence of both temporal boundaries~\cite{Aharonov_prl_1988}. The intrinsic tunability and conceptual versatility of WVs have led to broad applications in various quantum information and foundational studies~\cite{Ritchie_prl_1991,Lundeen&s_PhysRevLett_09,lundeen_nature_2011,Sacha_science_2011,Pusey_PhysRevLett_2014,xl_PhysRevLett_2024, Cao_PR_2026}. In particular, anomalous WVs that lie outside the eigenvalue spectrum of the measured observables have attracted considerable interest in precision metrology for their practical advantages in amplifying ultra-small physical effects~\cite{Duck_prd_1989, PaulKwiat_Science_2008, Dixon_PhysRevLett_2009, Steinberg_nature_2010, Brunner_prl_2010, Howell_PhysRevA_2010, Feizpour_prl_2011, Bruder_prl_2013, Boyd_prl_2014, Hallaji2017, Qu2020}. 

Despite these successes, the WV formalism still faces an important operational limitation. In the conventional setting, the pre-selected state is deterministically prepared and fixed in advance, which prevents branch-dependent access to distinct effective pre-/post-selected pairs. Consequently, anomalous WVs can only emerge from the branch corresponding to a ``successful" post-selection, whose occurrence is inherently probabilistic and typically suppressed~\cite{Dressel_RevModPhys_2014,XU_pqe_2024}. This constraint not only limits their operational accessibility, but also introduces a structural asymmetry: post-selection naturally accommodates two complementary outcomes, while pre-selection remains singular and fixed. Such asymmetry challenges the time-symmetric spirit of the TSVF~\cite{Aharonov_PhysRevB_1964, Aharonov_2008}, raising questions about the completeness of its operational implementation.

Recent theoretical developments have explored the incorporation of chronology-violating structures into quantum information processing, particularly through simulations of closed timelike curves (CTCs) inspired by general relativity~\cite{Kurt_1949, lobo2010, wang_arxiv_2026}. In the Deutsch model, a fixed-point consistency condition leads to striking phenomena such as perfect discrimination of non-orthogonal states~\cite{Brun_PhysRevLett_2009, Bennett_prl_2009, Ringbauer2014}, quantum cloning~\cite{Brun_PhysRevLett_2013, Yuan_npj_2015}, and violations of uncertainty relations~\cite{Pienaar_prl_2013}. An alternative approach, the framework of post-selected CTCs (PCTCs), is often regarded as more physically grounded due to its compatibility with path-integral formulations~\cite{Lloyd_prd_2011, LloydSeth_prl_2011}. Conceptually linked to the Wheeler-Feynman absorber theory~\cite{Feynman_RevModPhys_1945}, PCTCs employ time-symmetric boundary constraints to model retrocausal effects and can be implemented using quantum teleportation~\cite{oreshkov_nphysics_2015, Price_2012, Leifer_Proceedings_2017,Wharton_RevModPhys_2020, Brun2012_fop, Bartkiewicz_PhysRevA_2019, Arvidsson2023, Song_PhysRevLett_2024, huang2025arxiv, song_2025_arxiv, ji2026}, enabling tasks such as optimal state transmission from future to past~\cite{Arvidsson2023} and agnostic parameter estimation~\cite{Song_PhysRevLett_2024}. These advances motivate the investigation of an operational route to a symmetric pre- and post-selection structure.

\begin{figure}[ht]
  \centering
  \includegraphics[width=0.95\linewidth]{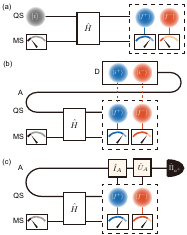}
  \caption{Schematic diagrams. (a) Standard WV measurement. (b) PCTC-powered WV measurement. (c) Post-selection-controlled WV measurement structure.}
  \label{circuit}
\end{figure}

In this work, we propose an experimentally accessible approach for realizing correlated pairwise TSVF by quantum entanglement and post-selection-controlled operations. In this architecture, post-selection is no longer merely a passive filtering step, but actively determines the corresponding pre-selected state. This enables the experimental observation of anomalous complementary WVs associated with mutually exclusive post-selection branches, an outcome that is inaccessible in the standard weak measurement setting. Our results not only provide an operational perspective on addressing a fundamental asymmetry in the TSVF, but also point to potential applications in post-selection-based precision metrology by improving event utilization across complementary branches.

\emph{Theoretical framework}.--We begin by introducing the standard WV measurement protocol, as depicted in Fig. \ref{circuit} (a). Consider a two-level QS initially pre-selected by $|i\rangle$, along with a meter state (MS) initialized in the state $|\Phi\rangle$. The QS and MS are weakly coupled via an interaction Hamiltonian $\hat{H} = g \delta (t-t_0) \hat{O}\otimes \hat{M}$, where $g \ll 1$ specifies the coupling strength, and $\hat{O}$ ($\hat{M}$) denotes the observable of the QS (MS). Following the interaction, the QS is post-selected onto a pair of orthogonal and complementary states $\{|f^+\rangle, |f^-\rangle\}$, which satisfy $\langle f^+|f^-\rangle = 0$ and $|f^+\rangle\langle f^+|+|f^-\rangle\langle f^-| = \hat{I}$. Conditioned on the complementary post-selection outcomes, the complementary WVs for a fixed pre-selected state $|i\rangle$ are given by
\begin{equation}
_{f^\pm}\langle \hat{O}\rangle_{i}^w = \frac{\langle f^\pm|\hat{O}|i\rangle}{\langle f^\pm|i\rangle}.
\end{equation}
An anomalous WV is characterized by a modulus exceeding the largest eigenvalue of the observable, e.g., $|_{f^+}\langle\hat{O}\rangle^w_{i}| > ||\hat{O}||_2$. This feature is exploited to amplify the weak interaction parameter $g$ in quantum sensing, known as weak-value amplification. The associated event is thus regarded as a ``successful" post-selection $|f^+\rangle$, occurring with probability $p_{f^+} \approx |\langle f^+|i\rangle|^2$. In contrast, the modulus of the WV $|_{f^-}\langle \hat{O}\rangle_{i}^w|$ post-selected by $|f^-\rangle$ remains confined within the eigenvalue spectrum~\cite{si}, and is typically discarded, leading to a significant loss of resources.

To enable simultaneous access to both anomalous complementary WVs, Arvidsson-Shukur et al. proposed a theoretical scheme based on PCTCs, as illustrated in Fig. \ref{circuit} (b). In this framework, the PCTCs are simulated probabilistically via a quantum teleportation protocol. The symbol $\subset$ denotes the preparation of a maximally entangled Bell state $|\Psi\rangle_{SA}$, shared between the QS and an ancilla A. The QS then undergoes a weak measurement followed by post-selection. Upon obtaining a post-selection outcome $|f^+\rangle$ or $|f^-\rangle$, the result is classically communicated to an ancilla D, which prepares the corresponding quantum state $|i^+\rangle_D$ or $|i^-\rangle_D$. The joint state of A and D is then projected onto the Bell state $|\Psi\rangle_{AD}$ (denoted by $\supset$), effectively teleportating the state $|i^\pm\rangle_D$ from the future back to the pre-selection of the QS. This CTC-based process facilitates access to anomalous complementary WVs $_{f^\pm}\langle\hat{O}\rangle_{i^\pm}^w$, with correlated pairwise pre- and post-selected states (i.e., $\langle f^+|\ |i^+\rangle$ and $\langle f^-|\ |i^-\rangle$).

\begin{figure*}[ht]
  \centering
\includegraphics[width=0.95\linewidth]{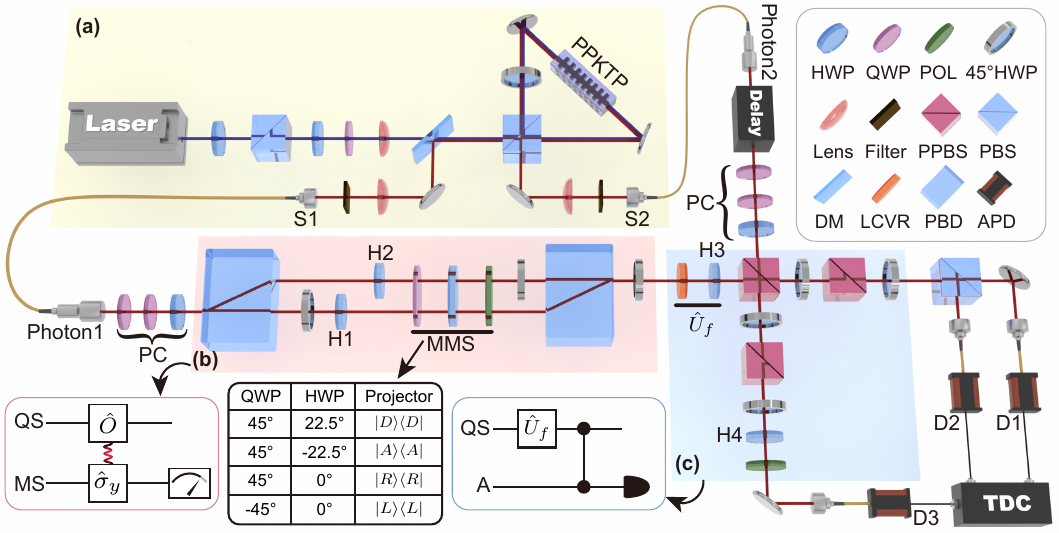}
  \caption{Experimental Setup. (a) A continuous-wave laser at 405 nm pumps a periodically poled $\text{KTiOPO}_4$ (PPKTP) crystal in the Sagnac interferometer, enabling spontaneous parametric down-conversion to prepare a two-photon polarization-entangled state. (b) Measurement of the WVs. (c) The post-selection-controlled structure, composed of $\hat{U}_f$, a controlled-Z gate followed by a projective measurement on the ancilla. Abbreviations: DM, dichroic mirror; APD, avalanche photodiode; TDC, time-to-digital converter.}
  \label{exp}
\end{figure*}

We introduce a post-selection-controlled structure built upon the WV framework, incorporating key conceptual elements from CTC-based approaches, as illustrated in Fig. \ref{circuit} (c). In the preparation stage, the QS and ancilla are initialized in a maximally entangled state:
\begin{equation}
|\Psi_i\rangle_{SA}=\frac{1}{\sqrt{2}} \left( |i^+\rangle_S|a^+\rangle_A + |i^-\rangle_S|a^-\rangle_A \right),
\label{eq2}
\end{equation}
where $\{|i^\pm\rangle\}$ and $\{|a^\pm\rangle\}$ are orthogonal basis states for the system and ancilla, respectively. The post-selection-controlled framework is implemented by applying a controlled unitary operation $\hat{U}_{SA} = |f^+\rangle\langle f^+|\otimes \hat{I}_A + |f^-\rangle\langle f^-|\otimes \hat{U}_A$, which acts on the ancilla conditioned on the post-selected state $|f^\pm\rangle$ of the QS, where the ancilla operation is defined by $\hat{U}_A|a^+\rangle = |a^-\rangle$. After the controlled unitary, the ancilla is projected onto $\Hat{\Pi}_{a^+}={|a^+\rangle\langle a^+|}$. Together with the initial entangled state, this projection and the controlled operation establish a one-to-one correlation between the post-selected state of the QS and the corresponding pre-selected state. Specifically, depending on the post-selection outcome, $f^{+}$ or $f^{-}$, this leads to the correlated pairwise two-state vectors $\langle f^{+}||i^{+}\rangle$ or $\langle f^{-}||i^{-}\rangle$, providing an operational realization of a time-symmetric structure. Within this framework, the denominators of the complementary WVs, i.e., $\langle f^+|i^+\rangle$ and $\langle f^-|i^-\rangle$, can be made to approach 0, allowing the simultaneous emergence of anomalous WVs. Moreover, when the observable $\hat{O}$ satisfies appropriate conditions, the complementary WVs have equal magnitude~\cite{si}.

To extract the WV, we initialize a qubit MS in the state $|\Phi\rangle = |0\rangle$. The interaction between the QS and the MS is governed by a Hamiltonian involving observables satisfying $\hat{O}^2=\hat{I}$ and $\hat{M}=\hat{\sigma}_y$, where $\hat{\sigma}_{x,y,z}$ denote the standard Pauli operators. Under these conditions, the system-meter coupling is described by the unitary operator $\hat{U}_c = \text{exp}(-ig\hat{O}\otimes \hat{\sigma}_y) = \cos g\hat{I}_S\otimes \hat{I}_M - i\sin g\hat{O}\otimes \hat{\sigma}_y$. Following this interaction, the QS is post-selected onto the complementary states $|f^\pm\rangle$, resulting in the corresponding final MSs $|\Phi^\pm\rangle$. The real and imaginary parts of the complementary WVs can then be extracted by measuring the observables $\hat{\sigma}_x$ and $\hat{\sigma}_y$ on $|\Phi^\pm\rangle$, yielding
\begin{equation}
_{f^\pm}\langle\hat{O}\rangle^w_{i^\pm} = \frac{\mathcal{N}}{\sin(2g)} (\left\langle\hat{\sigma}_{x}\right\rangle_{M}^\pm + i\left\langle\hat{\sigma}_{y}\right\rangle_{M}^\pm),
\end{equation}
where $\langle \hat{\sigma}_{x,y} \rangle_M^\pm$ are the expectation values conditioned on the post-selection outcome $f^\pm$ and the coefficient $\mathcal{N}=\cos ^{2}(g)+\sin ^{2}(g)|_{f^\pm}\langle \hat{O}\rangle^w_{i^\pm}|^{2}$.

We analyze the specific case implemented in our experiment with a traceless observable $\hat{O} = -\hat{\sigma}_z$. The pre- and post-selected states are parameterized as $|i^+\rangle = \cos\theta_i|0\rangle +\sin\theta_ie^{i\phi_i}|1\rangle$ and $|f^+\rangle = \cos\theta_f|0\rangle +\sin\theta_fe^{i\phi_f}|1\rangle$, respectively. When $\phi_i-\phi_f= 0$ and $\theta_i-\theta_f\to\frac{\pi}{2}$, the complementary WVs become real and anomalous, given by $_{f^+}\langle\hat{O}\rangle_{i^+}^w=-_{f^-}\langle\hat{O}\rangle_{i^-}^w = -\cos\left(\theta_i+\theta_f\right)/\cos\left(\theta_i-\theta_f\right)$. Alternatively, when $\theta_i+\theta_f =\frac{\pi}{2}$ and $\phi_i-\phi_f\to\pi$, the complementary WVs become purely imaginary and anomalous, with $_{f^+}\langle\hat{O}\rangle_{i^+}^w=_{f^-}\langle\hat{O}\rangle_{i^-}^w =i\tan[(\phi_i-\phi_f)/2]$.
\begin{figure*}[ht]
  \centering
  \includegraphics[width=\textwidth]{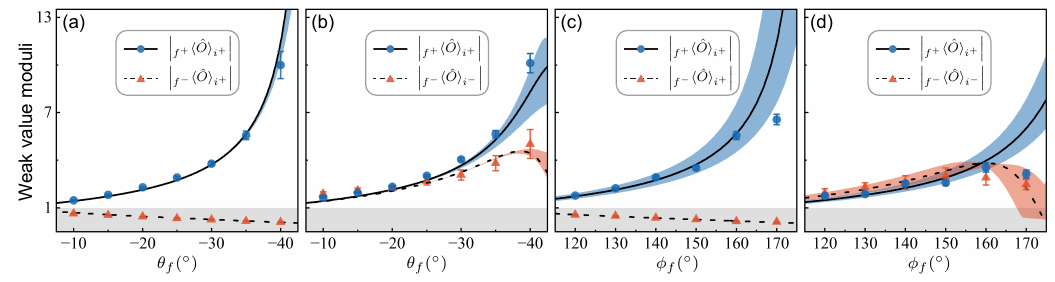}
  \caption{Experimentally measured moduli of complementary WVs. (a, b) Real WVs in the standard and post-selection-controlled protocols. (c, d) Imaginary WVs in the standard and post-selection-controlled protocols. Solid and dashed curves show theoretical predictions. Colored regions represent simulated errors assuming a phase instability of $\pm 5^\circ$ in the PBD interferometer. Blue circles and red triangles indicate moduli of WVs post-selected on $|f^+\rangle$ and $|f^-\rangle$, respectively. Error bars represent one standard deviation derived from Poissonian counting statistics and are smaller than the markers when not visible.}
  \label{data}
\end{figure*}

\emph{Experiment}.--In the experiment, we implement the post-selection-controlled protocol on a photonic platform, as illustrated in Fig. \ref{exp}. In module (a), a two-photon polarization-entangled state $|\Psi\rangle_{SA} = \frac{1}{\sqrt{2}}(|H\rangle_1|V\rangle_2 - |V\rangle_1|H\rangle_2)$ is generated. We refer to the polarization degree of freedom of photons 1 and 2 as the QS and ancilla, respectively. Following Eq. \eqref{eq2}, we assign $|i^\pm\rangle = (|H\rangle\pm|V\rangle)/\sqrt{2}$ and $|a^\pm\rangle = (|H\rangle\mp|V\rangle)/\sqrt{2}$. At the output ports of the fibers, each photon passes through two quarter-wave plates (QWPs) and one half-wave plate (HWP) for polarization compensation (PC)~\cite{wsk2007_cpl}.

Figure \ref{exp} (b) illustrates the experimental setup for WV measurement. First, a polarizing beam displacer (PBD) maps the polarization-encoded QS to spatial modes: $|H\rangle \rightarrow |0\rangle$ (upper path) and $|V\rangle \rightarrow |1\rangle$ (lower path). A HWP at $45^\circ$ in path 1 initializes both paths in the same polarization state $|H\rangle$, which serves as the MS. The QS observable is taken to be $\hat{O} = -\hat{\sigma}_z$. The coupling Hamiltonian $\hat{H} = -g\hat{\sigma}_z\otimes \hat{\sigma}_y$ is realized by rotating H1 (in path 1) to ${g}/{2}$ degrees and H2 (in path 0) to $-{g}/{2}$ degrees. Since the measurements on the MS (MMS) and post-selection of the QS act on separate Hilbert spaces, they commute and can be performed independently. By fixing all polarizers (POLs) at $0^\circ$, MMS are performed first, using a QWP-HWP-POL combination. The projectors associated with various wave plate settings are summarized in the accompanying table. Subsequently, a HWP at $45^\circ$ in path 0 and the second PBD recombine the spatial modes, implementing the map $\{|0\rangle\rightarrow |V\rangle, |1\rangle\rightarrow |H\rangle\}$. The final HWP at $45^\circ$ then applies $\hat{\sigma}_x$ operation, restoring the polarization-encoded QS for subsequent operation and measurement.

The experimental implementation of the post-selection-controlled operation is illustrated in Fig. \ref{exp} (c). For photon 1, a liquid crystal variable retarder (LCVR) and a HWP (H3) implement the unitary $\hat{U}_f$. The LCVR introduces a tunable relative phase $-\phi_f$ between $|H\rangle$ and $|V\rangle$ and H3 is set at $\theta_f/2$, transforming the basis from $\{|f^+\rangle,|f^-\rangle\}$ to $\{|H\rangle, |V\rangle\}$. Finally, a polarizing beam splitter (PBS) completes the post-selection on the QS, routing the $|f^+\rangle$ state to detector D1 and the $|f^-\rangle$ state to detector D2. The unitary required to transform $|a^-\rangle$ to $|a^+\rangle$ is $\hat{U}_A=\hat{\sigma}_z$. Accordingly, the controlled operation $\hat{U}_{SA}$ takes the form of a controlled-Z (CZ) gate $\hat{U}_{\text{CZ}}=|H\rangle_S\langle H|\otimes \hat{I}_A + |V\rangle_S\langle V|\otimes \hat{\sigma}_{z,A}$. This CZ gate is realized via two-photon Hong-Ou-Mandel interference on a partially polarizing beam splitter (PPBS)~\cite{Kiesel_prl_2005}. The PPBS is designed such that photons in $|H\rangle$ are fully transmitted, while photons in $|V\rangle$ are transmitted and reflected with probabilities 1/3 and 2/3, respectively. Two additional PPBSs along with four HWPs fixed at $45^\circ$ are used to balance the horizontal polarization. After the quantum-controlled operation, photon 2 is projected onto $|a^+\rangle$ using an HWP H4 set at $-22.5^\circ$ and a POL, with the transmitted photons detected at D3.

As a comparison, in the standard WV measurement, photons 1 and 2 are prepared in separable polarization states. Photon 2 serves as a herald, while photon 1 is initialized in a fixed pre-selected state using a POL and an HWP. Analogous to the setup shown in Fig. \ref{exp}, the post-selection procedure is implemented by $\hat{U}_f$ and the final PBS~\cite{si}.

In Fig. \ref{data}, we compare the moduli of WVs obtained using the standard protocol and through our post-selection-controlled framework. Panels (a) and (b) correspond to the purely real WVs, realized by setting $\theta_i = 45^\circ$, $\phi_i = \phi_f = 0^\circ$, and varying $\theta_f$ from $-10^\circ$ to $-40^\circ$. Panel (a) shows the results from the standard protocol. Here, the anomalous WVs with their moduli $|{}_{f^+}\langle\hat{O}\rangle_{i^+}|=\tan(45^\circ -\theta_f)$ are observed only when the post-selection outcome is $f^+$. In contrast, the complementary outcome $f^-$ yields moduli of the form $|{}_{f^-}\langle\hat{O}\rangle_{i^+}| =\cot(45^\circ -\theta_f)$, which remain bounded within the eigenvalue spectrum of the observable (indicated by the gray shaded region). Panel (b) presents the results from the post-selection-controlled protocol. In this case, the moduli of both complementary WVs $|{}_{f^\pm}\langle\hat{O}\rangle_{i^\pm}|=\tan(45^\circ - \theta_f )$ simultaneously exceed the eigenvalue bounds, enabled by the correlation of the pre/post-selection states. Notably, at $\theta_f=-40^\circ$, the post-selection-controlled protocol yields a maximum WV modulus of $|_{f^-}\langle \hat{O}\rangle_-| = 5.02 $, in stark contrast to the standard protocol, which produces only $|_{f^-}\langle \hat{O}\rangle_+| = 0.099$ with the same post-selected state $|f^-\rangle$.

Fig. \ref{data} (c) and (d) compare the moduli of imaginary WVs obtained using the standard and post-selection-controlled protocols, respectively. This regime is accessed by setting $\theta_i = \theta_f = 45^\circ$, $\phi_i = 0^\circ$, and varying $\phi_f$ from $120^\circ$ to $170^\circ$. In the standard protocol, anomalous WVs $|{}_{f^+}\langle\hat{O}\rangle_{i^+}|=\tan(\phi_f/2)$ appear only for the post-selection outcome $f^+$, while the complementary WVs corresponding to $f^-$, given by $|{}_{f^-}\langle\hat{O}\rangle_{i^+}|=\cot(\phi_f/2)$ remain within eigenvalue spectrum. In contrast, the post-selection-controlled scheme yields anomalous WVs $|{}_{f^\pm}\langle\hat{O}\rangle_{i^\pm}|=\tan(\phi_f/2)$ for both post-selection outcomes. Specifically, at $\phi_f=150^\circ$, the post-selection-controlled protocol achieves a maximum WV modulus of $|_{f^-}\langle \hat{O}\rangle_-| = 3.01$, significantly surpassing the standard protocol's value of $|_{f^-}\langle \hat{O}\rangle_+| = 0.27$.

%operationally accessible, extenability, while preserving....

\emph{Discussion and conclusions}.--In the standard protocol, the pre- and post-selected states are well controlled, allowing the theoretical curves in Fig. \ref{data} (a) and (c) to be calculated under idealized assumptions. In contrast, for the post-selection-controlled protocol, imperfections in entangled state generation necessitate the use of experimentally reconstructed states to derive the theoretical curves for $|{}_{f^\pm}\langle\hat{O}\rangle_{i^\pm}|$ in Fig. \ref{data} (b) and (d), leading to deviations from $|{}_{f^+}\langle\hat{O}\rangle_{i^+}|$ in panels (a) and (c). Discrepancies between the experimental data and theory primarily stem from statistical uncertainties, phase fluctuations in the interferometric paths, and non-ideal performance of the CZ gate. The larger error regions associated with the imaginary WVs in Fig. \ref{data} (c) and (d), compared to the real ones in panels (a) and (b), indicate that the imaginary WVs are more susceptible to phase fluctuations. Despite these experimental imperfections, our results clearly demonstrate the simultaneous observation of anomalous complementary WVs with $|{}_{f^\pm}\langle\hat{O}\rangle_{i^\pm}| \gg ||\hat{O}||_2$ enabled by the post-selection-controlled framework.

Compared with the proposal of Arvidsson-Shukur et al.~\cite{Arvidsson2023}, both schemes exploit quantum entanglement. Our scheme replaces the feed-forward quantum teleportation with post-selection–controlled operations within the entangled architecture. This preserves the key operational feature underlying the metrological advantage of anomalous WVs: the effective pre-selected state that creates an anomalous WV can be designated according to the subsequent weak interaction and the post-selection, instead of being fixed a priori. In this sense, our approach is closely aligned with the emerging notion of agnostic metrology, where the optimal probe need not be specified before the sensing interaction~\cite{Song_PhysRevLett_2024}.

Beyond reproducing the operational effect of the PCTC-assisted scheme, our protocol offers two practical advantages in terms of efficiency and extensibility. First, by eliminating the additional Bell measurements in teleportation-based implementations, we double the overall success probability, thereby improving event utilization~\cite{VAIDMAN_mpb_2006, Collins_quantum_2024, si}. Second, our experimental architecture provides a versatile platform for studying generalized TSVF. In our setting, correlations between the paired two-state vectors are classically heralded by the post-selection outcome. The setup can be straightforwardly extended to generate coherent superpositions of distinct two-state vectors by projecting onto a superposed post-selection basis. Moreover, our implementation is compatible with entanglement swapping~\cite{Aharonov_2008}, opening a route to exploring superpositions of causal order within pre- and post-selection frameworks~\cite{Onur_arxiv_2024, Nowakowski_pra_2018}.

In summary, we theoretically propose and experimentally implement correlated pairwise two-state vectors via quantum entanglement and post-selection-controlled operations. This experimentally accessible protocol addresses the intrinsic asymmetry in the realization of TSVF through pre- and post-selection, thus providing a concrete route toward richer TSVF structures. In contrast to standard WV measurements, our protocol elevates post-selection from a passive filtering step to an active conditioning mechanism, such that each post-selection outcome is paired with its own effective pre-selected state. Our results thus pave the way for experimentally exploring time-symmetric quantum descriptions and their applications in quantum metrology.

% \section*{Acknowledgments}
\emph{Acknowledgements}.--This work was supported by Quantum Science and Technology-National Science and Technology Major Project (Grant No. 2024ZD0300900), the National Natural Science Foundation of China (Grant Nos. 12347104, U24A2017, 12461160276), National Key Research and Development Program of China (Grant No. 2023YFC2205802), Natural Science Foundation of Jiangsu Province (Grants Nos. BK20243060, BK20233001), in part by State Key Laboratory of Advanced Optical Communication Systems and Networks, China. L. X. was funded by the National Natural Science Foundation of China (Grant Nos. 12305034, 92576111) and Basic Research Program of Jiangsu (Grant No. BK20251994).

\emph{Data availability}.--The data that support the findings of this article are openly available~\cite{xie_2026}.
\makeatletter
\let\arxivSavedFMNList\@FMN@list
\makeatother

\clearpage
\onecolumngrid
\appendix
\clearpage
\appendix
% --- Supplementary Material Settings ---

% 1. 重置计数器
\setcounter{figure}{0}
\setcounter{table}{0}
\setcounter{equation}{0}
\setcounter{page}{1} % 如果你想让页码也从 1 开始重新计数（可选）

% 2. 重新定义编号格式
\renewcommand{\thefigure}{S\arabic{figure}}
\renewcommand{\thetable}{S\arabic{table}}
\renewcommand{\theequation}{S\arabic{equation}}

\section{Supplementary Material}
\subsection{Standard protocol cannot simultaneously yield anomalous weak values for two complementary post-selections}

In this section, we show that a standard weak values (WVs) measurement protocol with a fixed pre-selected state cannot yield anomalous WVs in both complementary post-selection branches. We consider a general pre-selected state described by a density matrix $\rho_i$, satisfying $\rho_i\geq 0$ and $\operatorname{Tr}(\rho_i)=1$.

Let $\{|f^+\rangle,|f^-\rangle\}$ be two orthogonal and complete post-selection states, with projectors
\begin{equation}
\hat{\Pi}_{f^\pm}=|{f^\pm}\rangle\langle {f^\pm}|,
\end{equation}
satisfying
\begin{equation}
\hat{\Pi}_{f^+}+\hat{\Pi}_{f^-}=\mathds{1}.
\end{equation}
For a Hermitian observable $\hat O$ with spectral norm $\|\hat O\|_2$, the WVs associated with the two post-selections are defined as
\begin{equation}
_{f^\pm}\langle\hat O\rangle^w_{i}
=
\frac{
\operatorname{Tr}\left(\hat{\Pi}_{f^\pm}\hat O\rho_i\right)
}{
\operatorname{Tr}\left(\hat{\Pi}_{f^\pm}\rho_i\right)
},
\end{equation}
with post-selection probabilities
\begin{equation}
p_\pm
=
\operatorname{Tr}\left(\hat{\Pi}_{f^\pm}\rho_i\right).
\end{equation}
We assume $p_\pm\neq 0$ so that both WVs are well defined. For each branch, we have
\begin{equation}
p_\pm
\left|
_{f^\pm}\langle\hat O\rangle^w_{i}
\right|^2
=
\frac{
\left|
\operatorname{Tr}\left(\hat{\Pi}_{f^\pm}\hat O\rho_i\right)
\right|^2
}{
p_\pm
}=
\frac{
\left|
\langle f^\pm|\hat O\rho_i|f^\pm\rangle
\right|^2
}{
p_\pm
}.
\end{equation}
Using the Cauchy-Schwarz inequality for the positive semidefinite inner product
\begin{equation}
\langle x,y\rangle_{\rho_i}
=
\langle x|\rho_i|y\rangle,
\end{equation}
with $|x\rangle=\hat O|f^\pm\rangle$ and $|y\rangle=|f^\pm\rangle$, we obtain
\begin{equation}
\left|
\langle f^\pm|\hat O\rho_i|f^\pm\rangle
\right|^2
\leq
\langle f^\pm|\hat O\rho_i\hat O|f^\pm\rangle
\langle f^\pm|\rho_i|f^\pm\rangle.
\end{equation}
Equivalently,
\begin{equation}
\left|
\operatorname{Tr}\left(\hat{\Pi}_{f^\pm}\hat O\rho_i\right)
\right|^2
\leq
\operatorname{Tr}\left(\hat{\Pi}_{f^\pm}\hat O\rho_i\hat O\right)
\operatorname{Tr}\left(\hat{\Pi}_{f^\pm}\rho_i\right).
\end{equation}
Therefore,
\begin{equation}
p_\pm
\left|
_{f^\pm}\langle\hat O\rangle^w_{i}
\right|^2
\leq
\operatorname{Tr}\left(\hat{\Pi}_{f^\pm}\hat O\rho_i\hat O\right).
\end{equation}

Summing over the two complementary post-selection branches and using the completeness relation, we obtain
\begin{align}
p_+
\left|
_{f^+}\langle\hat O\rangle^w_{i}
\right|^2
+
p_-
\left|
_{f^-}\langle\hat O\rangle^w_{i}
\right|^2
&\leq
\operatorname{Tr}\left(\hat{\Pi}_{f^+}\hat O\rho_i\hat O\right)
+
\operatorname{Tr}\left(\hat{\Pi}_{f^-}\hat O\rho_i\hat O\right)
\\
&=
\operatorname{Tr}\left[
\left(
\hat{\Pi}_{f^+}+\hat{\Pi}_{f^-}
\right)
\hat O\rho_i\hat O
\right]
\\
&=
\operatorname{Tr}\left(\hat O\rho_i\hat O\right)
\\
&=
\operatorname{Tr}\left(\rho_i\hat O^2\right)
\\
&\leq
\|\hat O\|_2^2.
\end{align}
Therefore, if one branch is anomalous in the sense that
\begin{equation}
\left|
_{f^+}\langle\hat O\rangle^w_{i}
\right|
>
\|\hat O\|_2,
\end{equation}
then
\begin{align}
p_-
\left|
_{f^-}\langle\hat O\rangle^w_{i}
\right|^2
&\leq
\|\hat O\|_2^2
-
p_+
\left|
_{f^+}\langle\hat O\rangle^w_{i}
\right|^2
\\
&<
\|\hat O\|_2^2
-
p_+\|\hat O\|_2^2
\\
&=
p_-\|\hat O\|_2^2.
\end{align}
Since $p_->0$, this implies
\begin{equation}
\left|
_{f^-}\langle\hat O\rangle^w_{i}
\right|
<
\|\hat O\|_2.
\end{equation}
Thus, in a standard WVs measurement protocol with a fixed pre-selected state, two complementary post-selection branches cannot both yield WVs whose moduli exceed the spectral norm of the measured observable.
\subsection{Conditions for the equality of complementary WVs}
Achieving equal magnitudes for the complementary WVs with pre-selected state $|i^\pm\rangle$ with projectors $\hat{\Pi}_{i^\pm} = |{i^\pm}\rangle\langle {i^\pm}|(\hat{\Pi}_{i^+}+\hat{\Pi}_{i^-} =\mathds{1} )$ and the corresponding post-selected state $|f^\pm\rangle$ requires that 
\begin{equation}
    \left|\frac{\langle f^+| \hat{O} |i^+\rangle }{\langle f^+|i^+\rangle}\right| = \left|\frac{\langle f^-| \hat{O} |i^-\rangle }{\langle f^-|i^-\rangle}\right|\longrightarrow \operatorname{Tr}\left(\hat{\Pi}_{f^+}\hat{O}\hat{\Pi}_{i^+}\hat{O}\right) = \operatorname{Tr}\left(\hat{\Pi}_{f^-}\hat{O}\hat{\Pi}_{i^-}\hat{O}\right)  = \operatorname{Tr}\left[\left(\mathds{1}-\hat{\Pi}_{f^+}\right)\hat{O}\left(\mathds{1}-\hat{\Pi}_{i^+}\right)\hat{O}\right],
\end{equation}
which indicates that both the pre- and post-selected states, along with the observable, must fulfill the condition
\begin{equation}
\operatorname{Tr}\left(\hat{O}^2\right) = \operatorname{Tr}\left[\left(\hat{\Pi}_{i^+} + \hat{\Pi}_{f^+}\right)\hat{O}^2\right].
\label{condition}
\end{equation}
Consider an arbitrary two-dimensional Hermitian operator $\hat{O}$, which can be decomposed into the identity and Pauli components
\begin{equation}
    \hat{O} = a_0 \mathds{1} + \vec{a} \cdot \vec{\sigma},
\end{equation}
where $a_0 \in \mathbb{R}$, $\vec{a} \in \mathbb{R}^3$, and $\vec{\sigma} = (\sigma_x, \sigma_y, \sigma_z)$ is the vector of Pauli matrices. Using the identity $(\vec{a} \cdot \vec{\sigma})^2 = |\vec{a}|^2 \mathds{1}$, the square of the observable $\hat{O}$ can be calculated as $\hat{O}^2 = (a_0^2 + |\vec{a}|^2)\mathds{1} + 2a_0 (\vec{a} \cdot \vec{\sigma})
$. We also define the pre-selected and post-selected states using rank-1 projectors, represented on the Bloch sphere as 
\begin{equation}
    \hat{\Pi}_{i^+} = \frac{\mathds{1} + \vec{r}_i \cdot \vec{\sigma}}{2}, \quad \hat{\Pi}_{f^+} = \frac{\mathds{1} + \vec{r}_f \cdot \vec{\sigma}}{2},
\end{equation}
where $\vec{r}_i$ and $\vec{r}_f$ are real unit vectors ($|\vec{r}_{i,f}| = 1$). 

We now substitute these expressions into the condition Eq. \ref{condition}. The left-hand side (LHS) evaluates to
\begin{equation}
    \text{LHS} = \text{Tr}\left[ (a_0^2 + |\vec{a}|^2)\mathds{1} + 2a_0 (\vec{a} \cdot \vec{\sigma}) \right] = 2(a_0^2 + |\vec{a}|^2).
\end{equation}
For the right-hand side (RHS), the trace calculation yields
\begin{equation}
    \begin{aligned}
\text{RHS} &= \text{Tr}\left[ \left(\mathds{1} + \frac{\vec{r}_i + \vec{r}_f}{2} \cdot \vec{\sigma}\right) \left( (a_0^2 + |\vec{a}|^2)\mathds{1} + 2a_0 (\vec{a} \cdot \vec{\sigma}) \right) \right] \\
&= \text{Tr}(\hat{O}^2) + \text{Tr}\left[ \left( \frac{\vec{r}_i + \vec{r}_f}{2} \cdot \vec{\sigma} \right) \left( 2a_0 \vec{a} \cdot \vec{\sigma} \right) \right] \\
&= 2(a_0^2 + |\vec{a}|^2) + a_0 \text{Tr}\left[ (\vec{r}_i + \vec{r}_f) \cdot \vec{\sigma} (\vec{a} \cdot \vec{\sigma}) \right] \\
&= 2(a_0^2 + |\vec{a}|^2) + 2 a_0 \vec{a} \cdot (\vec{r}_i + \vec{r}_f).
\end{aligned}
\end{equation}
Equating LHS and RHS implies the concise necessary and sufficient condition
\begin{equation}
    a_0 \vec{a} \cdot (\vec{r}_i + \vec{r}_f) = 0.
    \label{s13}
\end{equation}
This condition holds under three specific scenarios: (a) $a_0 = 0$.
The operator is purely traceless ($\hat{O} = \vec{a} \cdot \vec{\sigma}$). In this case, $\hat{O}^2 = |\vec{a}|^2 \mathds{1}$, which is proportional to the identity. The condition is automatically satisfied regardless of the states $\hat{\Pi}_{i^+}$ and $\hat{\Pi}_{f^+}$. (b) $\vec{a} = \vec{0}$.
The operator is proportional to the identity ($\hat{O} = a_0 \mathds{1}$). Here, $\hat{O}^2 = a_0^2 \mathds{1}$, and the condition is trivially satisfied for any $\hat{\Pi}_{i^+}, \hat{\Pi}_{f^+}$. (c) For a general observable containing both identity and Pauli components, the condition requires a geometric orthogonality with $ \vec{a} \perp (\vec{r}_i + \vec{r}_f)$. 
\subsection{Preparation of entangled states and efficiency of CZ gate}
\begin{figure*}[ht]
  \centering
  \includegraphics[width=0.9\linewidth]{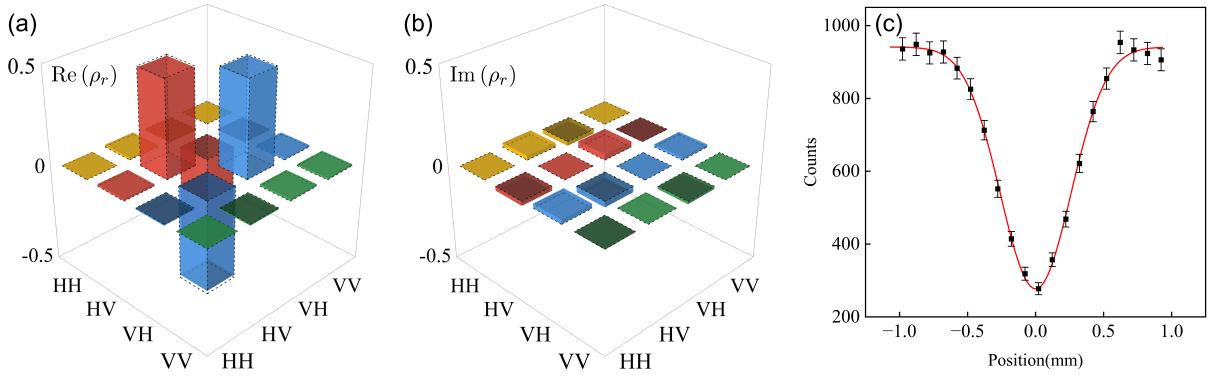}
  \caption{The density matrix representing the two-photon entangled state is depicted using bars and dashed edges, where the bars correspond to the experimentally reconstructed results, and the dashed edges denote the theoretical state. (a) The real part and (b) the imaginary part of the reconstructed state. (c) The Hong-Ou-Mandel Interference dip of the CZ gate with $|V\rangle$ photons injected into both ports.}
  \label{source}
\end{figure*}

Fig. 2 (a) in the main text illustrates the experimental configuration of the two-photon polarization-entangled source. The ideal target entangled state is $|\Psi^-\rangle = \frac{1}{\sqrt{2}}(|H\rangle|V\rangle - |V\rangle|H\rangle)$ and due to the experimental non-idealities, the results of the reconstructed state $\rho_r$ that characterized through quantum state tomography are presented in Fig. \ref{source} (a), (b). The fidelity between the ideal state $\rho_{id} = |\Psi^-\rangle \langle \Psi^-| $ and the reconstructed state $\rho_r$ is $ f\left(\rho_r, \rho_{id}\right) = \left( \mathrm{Tr}\sqrt{\sqrt{\rho_r} \rho_{id} \sqrt{\rho_r}}\right)^2=0.98\pm0.005$.

Fig. \ref{source} (c) is the Hong-Ou-Mandel Interference (HOMI) dip with visibility $\mathcal{V} = 0.705$. The theory visibility is $\mathcal{V}_{th} = 0.8$ because the PPBS we use has a transmission of 1 for H light and a transmission of 1/3 for V light. Conditioned on detecting one photon in each output, this CZ gate succeeds with a probability of $1/9$, which is an inherent limitation of the linear optical implementation rather than a constraint of our protocol, since there are deterministic ways to realize a CZ gate~\cite{SI:Heuck_PhysRevLett_2020, SI:krastanov_2022, SI:Alushi_PRXQuantum_2023}. The imperfection of the HOMI visibility stems from the continuous-wave pump laser and the PPBS crystal fabrication techniques.

\subsection{Experimental implementation of the standard WV measurement protocol}
\begin{figure*}[ht]
  \centering
  \includegraphics[width=\linewidth]{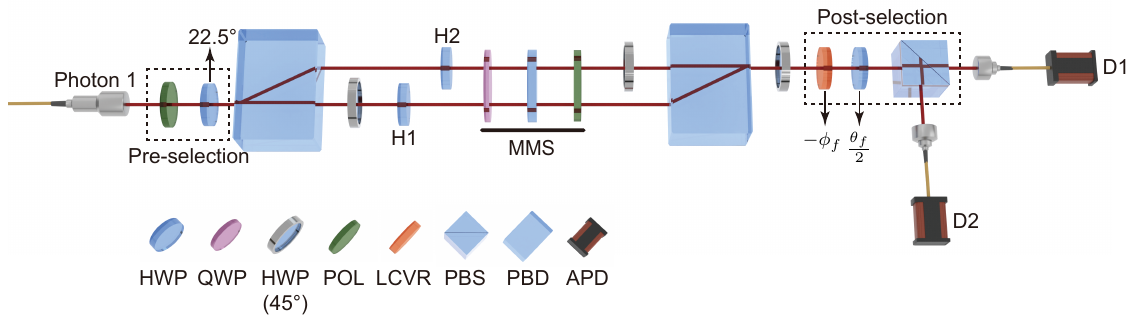}
  \caption{The standard WV measurement protocol experimental setup. The measurements of the meter state (MMS) are performed using a QWP-HWP-POL combination. Angles of H1, H2 and the wave plates in the MMS are consistent with the parameters described in the main text.}
  \label{standard_protocol}
\end{figure*}
The experimental setup for the standard weak value measurement protocol is illustrated in Fig. \ref{standard_protocol}. The setup consists of a pre-selection stage, a weak coupling and meter measurement module (identical to that shown in Fig. 2(b) of the main text), and a post-selection stage. The pre-selected state $|i^+\rangle = (|H\rangle+|V\rangle)/\sqrt{2}$ is prepared in the polarization degree of freedom of photon 1 using a HWP fixed at $22.5^\circ$. The angles of the two coupling HWPs follow the settings described in the main text. The post-selection stage is implemented via an LCVR-HWP-PBS combination, where detection by D1 (D2) corresponds to post-selection onto the state $|f^+\rangle$ ($|f^-\rangle$). Photon 2 serves as a heralding signal (not shown in the figure). The corresponding experimental results are presented in Fig. 3(a) and (c) of the main text.

\subsection{Success probability of different protocols}
While the success probability of extracting anomalous WVs assisted by PCTCs is typically reduced by half compared to the standard protocol $p_{s}$, and any attempt to double this probability often compromises the linkage with PCTCs~\cite{SI:Arvidsson2023}. The post-selection-controlled protocol offers a distinct advantage. In this section, we prove that the total success probability $p_h$ of the protocol recovers the standard value, i.e., $p_h = p_s$.

Consider a general coupling unitary $\hat{U}_{SM}$ between the QS and the MS. In the standard protocol, the unnormalized state of the MS after system post-selection is given by
\begin{equation}
    |\Tilde{\Phi}\rangle_u = \langle{f^+}|\hat{U}_{SM}|i^+\rangle|\Phi\rangle.
\end{equation}
The corresponding post-selected success probability is
\begin{equation}
    p_s={}_u \langle\Tilde{\Phi}|\Tilde{\Phi}\rangle_u = \operatorname{Tr}\left[\hat{U}_{SM}^\dagger \left(\hat{\Pi}_{f^+}\otimes \mathds{1}_M\right)\hat{U}_{SM}\left(\hat{\Pi}_{i^+}\otimes \hat{\Pi}_\Phi\right)\right],
\end{equation}
where $\hat{\Pi}_\Phi = |\Phi\rangle\langle \Phi|$ is the projector onto the initial MS state.

In the post-selection-controlled protocol, we introduce an auxiliary ancilla $A$ to implement the quantum-controlled operation $\hat{U}_{SA}$. The resulting unnormalized MS states, conditioned on the post-selected outcomes $f^\pm$ are
\begin{equation}
    |\Tilde{\Phi}\rangle^\pm_u = \langle{f^\pm}|\langle a^\pm|\hat{U}_{SA}\hat{U}_{SM}|\Psi\rangle_{SA}|\Phi\rangle.
\end{equation}
The post-selected success probabilities are given by
\begin{equation}
    p_h^\pm = {}^\pm_u \langle\Tilde{\Phi}|\Tilde{\Phi}\rangle^\pm_u = \frac{1}{2}\operatorname{Tr}\left[\hat{U}_{SM}^\dagger \left(\hat{\Pi}_{f^\pm}\otimes \mathds{1}_M\right)\hat{U}_{SM}\left(\hat{\Pi}_{i^\pm}\otimes \hat{\Pi}_\Phi\right)\right].
\end{equation}
To determine the total probability $p_h = p_h^+ + p_h^-$, we expand the expression for $2p_h^-$ by utilizing the completeness relation $\hat{\Pi}_{i^-/f^-} = \mathds{1}_S - \hat{\Pi}_{i^+/f^+}$
\begin{align}
    2p_h^-&= \operatorname{Tr}\left\{\hat{U}_{SM}^\dagger \left[\left(\mathds{1}_S-\hat{\Pi}_{f^+}\right)\otimes \mathds{1}_M\right]\hat{U}_{SM}\left[\left(\mathds{1}_S-\hat{\Pi}_{i^+}\right)\otimes \hat{\Pi}_\Phi\right]\right\}\\
    &=\operatorname{Tr}\left(\mathds{1}_S\otimes\hat{\Pi}_\Phi\right)-\operatorname{Tr}\left[\hat{U}_{SM}^\dagger \left(\hat{\Pi}_{f^+}\otimes \mathds{1}_M\right)\hat{U}_{SM}\left(\mathds{1}_S\otimes \hat{\Pi}_\Phi\right)\right]-\operatorname{Tr}\left(\hat{\Pi}_{i^+}\otimes\hat{\Pi}_\Phi\right)+2p_h^+\\
    & = 1-\operatorname{Tr}_S \left( \hat{\Pi}_{f^+} \cdot \hat{\rho}_S \right)+2p_h^+,
\end{align}
where the partial trace $\hat{\rho}_S = \operatorname{Tr}_M \left[ \hat{U}_{SM} \left(\mathds{1}_S\otimes \hat{\Pi}_\Phi\right) \hat{U}_{SM}^\dagger \right]$. For a von Neumann-type coupling
\begin{equation}
    \hat U_{SM}=e^{-ig\hat O\otimes\hat M}
    =\sum_n |n\rangle\langle n|\otimes \hat V_n,
\end{equation}
with $\hat O=\sum_n\lambda_n|n\rangle\langle n|$ and
$\hat V_n=e^{-ig\lambda_n\hat M}$, one finds
\begin{align}
    \hat{\rho}_S &= \operatorname{Tr}_M \left[ \left( \sum_n |n\rangle\langle n| \otimes \hat{V}_n \right) \left( \sum_k |k\rangle\langle k| \otimes \hat{\Pi}_\Phi \right) \left( \sum_m |m\rangle\langle m| \otimes \hat{V}_m^\dagger \right) \right]\\
    &= \sum_n |n\rangle\langle n| \cdot \operatorname{Tr}_M \left( \hat{V}_n \hat{\Pi}_\Phi \hat{V}_n^\dagger \right)\\
    &= \sum_n |n\rangle\langle n| = \mathds{1}_S.
\end{align}
We derive that for different post-selected outcomes, the success probabilities are same $p_h^- = p_h^+ = \frac{1}{2}p_s$, and we have
\begin{equation}
    p_h = p_h^+ + p_h^- = p_s.
\end{equation}

This difference originates from the underlying implementation. The PCTC-assisted scheme is simulated through post-selected teleportation, which introduces an additional probabilistic filtering step and hence an extra reduction in the overall success probability. By contrast, the post-selection-controlled protocol realizes the relevant alternatives coherently within a single controlled operation, so the success probability is consistent with the standard protocol.

\subsection{Real and imaginary parts of WVs}
\begin{figure*}[ht]
  \centering
  \includegraphics[width=0.9\linewidth]{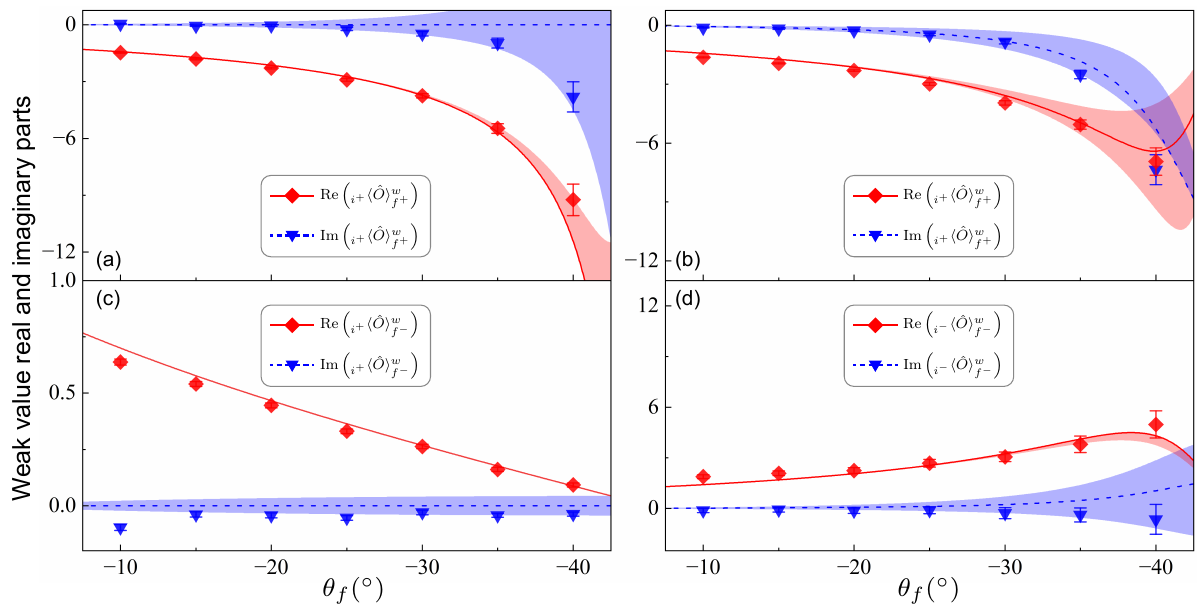}
  \caption{Real and imaginary parts of the WVs with $\theta_i = 45^\circ$, $\phi_i = \phi_f = 0^\circ$, where the theoretical WVs are purely real. Panels (a) and (c) show results from the standard WV measurement protocol with a fixed pre-selected state $|i^+\rangle$. Panels (b) and (d) show results obtained via our post-selection-controlled protocol. The solid and dashed lines represent theoretical predictions of the real and imaginary parts; for the quantum-controlled case, these calculations incorporate the realistic entangled state reconstructed via quantum state tomography. The shaded bands around the theory curves indicate simulated uncertainty assuming a phase instability of $\pm 5^\circ$ in the PBD interferometer. Error bars denote one standard deviation derived from Poissonian counting statistics; where not visible, they are smaller than the data points.}
  \label{real_data}
\end{figure*}

\begin{figure*}[ht]
  \centering
  \includegraphics[width=0.9\linewidth]{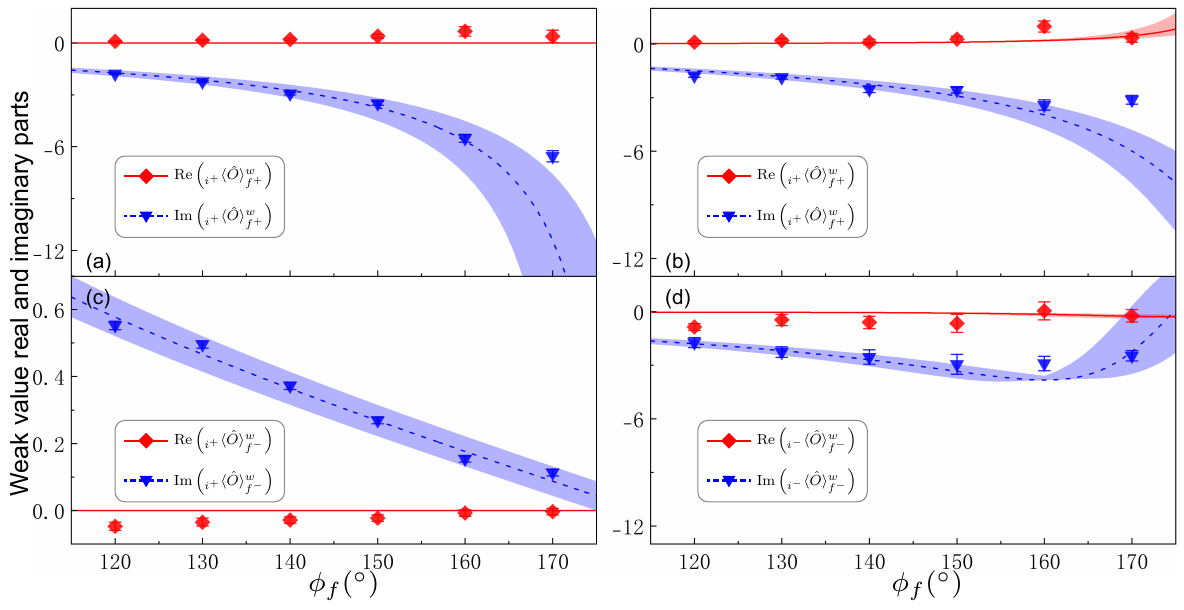}
  \caption{Real and imaginary parts of the WVs with $\theta_i = \theta_f = 45^\circ$, $\phi_i = 0^\circ$, where the theoretical WVs are purely imaginary. Panels (a) and (c) show results from the standard WV measurement protocol with a fixed pre-selected state $|i^+\rangle$. Panels (b) and (d) show results obtained via our post-selection-controlled protocol. The solid and dashed lines represent theoretical predictions of the real and imaginary parts; for the quantum-controlled case, these calculations incorporate the realistic entangled state reconstructed via quantum state tomography. The shaded bands around the theory curves indicate simulated uncertainty assuming a phase instability of $\pm 5^\circ$ in the PBD interferometer. Error bars denote one standard deviation derived from Poissonian counting statistics; where not visible, they are smaller than the data points.}
  \label{imag_data}
\end{figure*}

While the moduli of the WVs are presented in Fig. 3 of the main text, this section details their corresponding real and imaginary components. Fig. \ref{real_data} displays the results for the configuration $\theta_i = 45^\circ$ and $\phi_i = \phi_f = 0^\circ$, where the theoretical WVs are purely real. The figure is organized to facilitate comparison: the left column [panels (a), (c)] corresponds to the standard WV measurement protocol with a fixed pre-selected state $|i^+\rangle$, while the right column [panels (b), (d)] employs our post-selection-controlled protocol. The rows distinguish the post-selected states, with the upper row [(a), (b)] post-selected on $|f^+\rangle$ and the lower row [(c), (d)] on $|f^-\rangle$. Similarly, Fig. \ref{imag_data} presents the real and imaginary parts for the scenario $\theta_i = \theta_f = 45^\circ$, $\phi_i = 0^\circ$, in which the theoretical WVs are purely imaginary.

\begingroup
\makeatletter
\let\arxivOriginalLabel\label
\let\@FMN@list\arxivSavedFMNList
\def\NAT@bibsetnum#1{%
 \setlength{\topsep}{\z@}%
 \NATx@bibsetnum{4}%
}
\let\@bibsetup\NAT@bibsetnum
\def\present@bibnote#1#2{%
 \item[%
  \textsuperscript{%
   \normalfont
   \Hy@raisedlink{\hyper@anchorstart{SI.frontmatter.#1}\hyper@anchorend}%
   \begingroup
    \csname c@\@mpfn\endcsname#1\relax
    \frontmatter@thefootnote
   \endgroup
  }%
 ]#2\par
}
\renewcommand{\label}[1]{%
  \def\arxivCurrentLabel{#1}%
  \def\arxivLastBibItem{LastBibItem}%
  \ifx\arxivCurrentLabel\arxivLastBibItem
    \arxivOriginalLabel{LastSIBibItem}%
  \else
    \arxivOriginalLabel{#1}%
  \fi
}
\makeatother

\endgroup
\end{document}